\documentclass[prd,aps,twocolumn,amsmath,amssymb,nofootinbib,preprintnumbers]
{revtex4}

\usepackage{graphicx}% Include figure files
\usepackage{dcolumn}% Align table columns on decimal point
\usepackage{bm}% bold math
\usepackage{amsmath}
\usepackage{amsfonts}
\usepackage[caption=false]{subfig}
\usepackage{placeins}

\newcommand{\remove}[1]{}

\def\ls{\mathrel{\lower4pt\vbox{\lineskip=0pt\baselineskip=0pt
           \hbox{$<$}\hbox{$\sim$}}}}
\def\gs{\mathrel{\lower4pt\vbox{\lineskip=0pt\baselineskip=0pt
           \hbox{$>$}\hbox{$\sim$}}}}

\newcommand{\beq}{\begin{equation}}
\newcommand{\eeq}{\end{equation}}

\begin{document}

%
%\vspace*{2cm}
\title{Distinguishing among dark matter annihilation channels with neutrino telescopes}

\author{Rouzbeh Allahverdi$^{1}$}
% \author{Alexander Friedland$^{2}$}
\author{Katherine Richardson$^{1}$}

\affiliation{$^{1}$~Department of Physics \& Astronomy, University of New Mexico, Albuquerque, NM 87131, USA}
% \affiliation{$^{2}$~Theoretical Division, MS B285, Los Alamos National Laboratory, Los Alamos, NM 87545, USA}

%\date{April 13, 2011}

\begin{abstract}

We investigate the prospects for distinguishing dark matter annihilation channels using the neutrino flux from
gravitationally captured dark matter particles annihilating inside the sun.
We show that, even with experimental error in energy reconstruction taken into account, the spectrum of contained muon tracks may be used
to discriminate neutrino final states from the gauge boson/charged lepton final states and to determine their corresponding branching
ratios. We also discuss the effect of $\nu_\tau$ regeneration inside the sun as a novel method to distinguish the flavor of final state
neutrinos. This effect as evidenced in the muon spectrum becomes important for dark matter masses above 300 GeV. Distinguishing primary
neutrinos and their flavor may be achieved using multi-year data from a detector with the same capability and effective volume as the
IceCube/DeepCore array.

\medskip

%12.60.Jv = Supersymmetry
%98.80.Cq = Cosmology
\end{abstract}
%MIFP-09-25 \\ September 24, 2009
\maketitle
%%%%%%%%%%%%%%%%%%%%%%%%%%%%%%%%%%%%%%%%%%%%%%%%%%%%%%%%%%%%%%%%%%%%%%%%%%%%%%%
\section{Introduction}

Many lines of evidence support the existence of dark matter (DM) in the universe, but its identity remains a major problem
at the interface of particle physics and cosmology. The weakly interacting massive particles (WIMPs) are promising DM
candidates~\cite{WIMP}. They can explain the DM relic abundance, as precisely measured by cosmic microwave background (CMB)
experiments~\cite{WMAP7}, via thermal freeze-out of
annihilation in the early universe. Major experimental efforts seek to detect DM particles via direct, indirect and collider
searches. These experiments provide complementary information on the properties of DM such as its mass, elastic scattering cross
section with nucleons, annihilation cross section, and annihilation channels.

Indirect searches investigate annihilation of DM to various final states (neutrinos, photons, charged particles) through
astrophysical observations. Neutrinos provide an especially interesting probe because they are least affected on their way from the
production point to the detection point. As a result, neutrino telescopes like IceCube/DeepCore (IC/DC) can trace a neutrino signal
directly back to the source. DM particles gravitationally captured inside the sun annihilate and produce such a signal. When
equilibrium between DM capture by the sun and DM annihilation inside the sun is established, the flux of produced neutrinos depends on
the DM mass, scattering cross section off nucleons and the annihilation spectrum. The DM mass and scattering cross section can be
independently determined from other experiments. In particular, the Large Hadron Collider (LHC) is on the verge of discovering new physics,
which will enable us to measure the mass of the DM particle. Therefore, using the LHC measurements and the IC/DC results in tandem, we hope
to identify the annihilation channels and their corresponding branching ratios.

In this paper, we study the prospects for determining DM annihilation final states with IC/DC. This model-independent
study has two goals: 1) distinguishing scenarios in which DM annihilates into neutrinos from scenarios where DM annihilates into
gauge bosons and charged leptons, and 2) extracting information about flavor composition of neutrino final states in the former case.

Gauge boson and charged fermion final states are the dominant annihilation channels for the popular and extensively studied neutralino DM in
the minimal supersymmetric standard model (MSSM). On the other hand, annihilation to neutrinos can arise as the dominant channel in
extensions of the standard model (SM) in which DM is related to the neutrino sector. Discriminating neutrino final states from
gauge boson and tau final states will therefore allow us to discern between the two large classes of DM models. In the case that DM
mainly annihilates into neutrinos, it will be important to also know the flavor composition of the final state neutrinos. This will shed
further light on the specific aspects of the model that connect DM to neutrinos.

We show that the contained muon tracks at neutrino telescopes like IC/DC may be used for both purposes.
The spectrum of contained muons can be used to determine the branching ratios of neutrino final states versus gauge boson and tau final
states. We demonstrate this for several points in the parameter space within the reach of the one-year sensitivity
limits of the IC/DC array, with the background from atmospheric neutrinos and the experimental error in energy
reconstruction of
the muons taken into account. In addition, the $\nu_\tau$ regeneration due to charged current interactions inside the sun may be used to
extract information about the flavor of
final state neutrinos for DM masses above 300 GeV. In particular, we see that the $\nu_\tau$ final state can be discriminated from
the $\nu_\mu$ and $\nu_e$ final states at a significant level for DM masses as heavy as 800 GeV. Distinguishing the neutrino final
states and discriminating the flavor of neutrinos may be achieved with multi-year data from IC/DC.

This paper is organized as follows. The DM signal in neutrino telescopes is explained in section II. In section III, we discuss
the motivation for models where DM annihilates mainly into neutrinos. Section IV includes the analysis distinguishing the neutrino
final state from the gauge boson and tau final states. We investigate discriminating the neutrino flavor in the former case in section V.
Finally, we conclude the paper in Section VI.

%%%%%%%%%%%%%%%%%%%%
\section{Neutrino Telescopes as Dark Matter Detectors}
%%%%%%%%%%
\label{NeutrinoSignal}

\subsection{The Sun as a Source of Dark Matter Neutrinos}

DM annihilation in the sun produces a neutrino flux that can be probed by neutrino telescopes like IC/DC. This
neutrino flux is modeled by calculating the number of gravitationally captured DM particles in the sun and then considering the
propagation and detection of the neutrinos produced in DM annihilation. The number of
captured DM particles as a function of time is governed by a differential equation that balances the capture of particles from
elastic scattering on nucleons with the annihilation rate. The total rate of
annihilation in the sun is {\bf given by~\cite{Jungman:1995df} (also see~\cite{Baer:2004qq})}
\begin{equation} \label{annihilationeq}
\Gamma_A = \frac{C}{2} \tanh^2{\left ( \frac{t}{\tau_{eq}} \right )} \;,
\end{equation}
where $C$ is the capture rate of DM  particles. The number of particles captured will saturate at $\Gamma_A \approx C/2$ as
long as the length of time for the process has exceeded the equilibration time, $\tau_{eq} \equiv (\sqrt{C A})^{-1}$.

With a nominal freeze-out annihilation rate of $3 \times 10^{-26}$ $\mathrm{cm}^3/\mathrm{s}$, DM readily achieves equilibrium within the
lifetime of the solar system, $4.5$ Gyr, for spin-independent elastic scattering cross sections compatible with the limits set by the
XENON100
experiment: $\sigma_{\rm SI} = {\rm few} \times (10^{-9} - 10^{-8}) \; \mathrm{pb}$ for the mass range $100~{\rm
GeV}-1$ TeV~\cite{Aprile:2011hi}. The spin-dependent cross section needed to reach equilibrium is larger by a factor of about
$300$~\cite{Jungman:1995df}, which is well below the current experimental limit of $\sigma_{\mathrm{SD}} \leq 2.5 \times (10^{-4}-10^{-3})
\; \mathrm{pb}$ for the $100~{\rm GeV}-1 \; \mathrm{TeV}$ mass range~\cite{Collaboration:2011ec}.

The difference between the equilibrium constraints on $\sigma_{\rm SI}$ and $\sigma_{\rm SD}$ comes from the
fact that $\sigma_{\rm SI} \propto M^4$ {\bf while $\sigma_{SD} \propto J(J+1) M^2$}; $M$ and $J$ are the mass and spin of target nuclei inside
the sun respectively. Heavy elements such as iron account for only a thousandth of the sun's mass and suffer a
two order of magnitude form-factor suppression in capturing WIMP-scale DM masses~\cite{Jungman:1995df}. On the other hand, hydrogen
dominates the sun's composition and does not suffer a form factor suppression. Thus, the ratio of the spin-independent contribution from
iron to the spin-dependent contribution from hydrogen is on the order of $(56)^4 / 10^5 \sim {\cal O}(100)$. Since equilibrium is easily
achieved in the sun, the neutrino signal will depend solely on $C$, or equivalently
$\sigma_{\mathrm{SI}}$ and/or $\sigma_{\mathrm{SD}}$.\footnote{The neutrino signal from DM annihilation in the earth is
negligible assuming a standard DM velocity distribution in the halo.}

Once neutrinos are produced from DM annihilation in the sun, their spectra undergo a number of changes before
detection. Scattering via neutral current interactions results in neutrino energy loss, while charged current interactions
result in absorption of neutrinos. These effects are proportional to the energy of the neutrino and will therefore have the greatest effect
on the high energy part of the spectrum. Neutrinos are also affected by oscillations on the way to the earth.
Since the oscillation length is proportional to energy, $L_{\rm osc} \propto
E_\nu$, the effect is more important for the low energy part of the spectrum.

\subsection{Neutrino Background, Energy Reconstruction and Thresholds}

Neutrino telescopes access the neutrino signal from DM annihilation by recording Cerenkov light from relativistic charged particles
in their volume. Muon neutrinos produce muons via charged current interactions in the detector. Cosmic ray showers create a muon background
that can be controlled by selecting for upward-going events since muons are stopped in the earth. This limits observation of a DM
signal from the sun to half the year, when the sun is below the horizon.\footnote{In addition, a portion of the detector may be used as a
veto to observe contained muon events with a conversion vertex of $\nu_\mu$ to $\mu$ inside the instrumented volume, as in the case of
IC/DC. The veto procedure virtually eliminates the contribution to the background from through-going atmospheric muons by selecting for
contained vertices. This increases
the potential observation time to the full year when the sun is both above and below the horizon.} With through-going muons eliminated as a
background, the most significant contribution to the remaining background comes from atmospheric neutrinos. The spectrum of the atmospheric
neutrino background from cosmic rays is understood theoretically to within $20\%$~\cite{Honda} and is measured above 100 GeV to within
$10\%$~\cite{Abbasi:2010ie}, but individual atmospheric neutrino events cannot be distinguished from neutrinos from DM
annihilation.

The DM neutrino signal can be further enhanced over the background by allowing for an angular cut
in the direction of the sun. The cut will be made on the muon track, and not the incoming
neutrino. A smaller angle between the track and incident neutrino occurs for muons with an energy closer to that of the
neutrino. For an elastic collision, the dependence of the angle $\theta_{\nu \mu}$ on the muon energy $E_\mu$ angle is
approximately given by $\theta_{\nu \mu} \ls 5.7^\circ ~ (100~{\rm GeV}/E_\mu)^{1/2}$ (for example,
see~\cite{Cirelli}). That is, the muon events that deflect little from the incoming neutrino path are the highest energy events. While the
smallest possible angular cut is desirable to eliminate background, more accommodating cuts on the angle between the track and the sun
provide information about lower energy events.

Energy reconstruction of events can be approximated in two regimes. Above $1 \; \mathrm{TeV}$, the light generated by an event is
proportional to the muon energy since Bremsstrahlung radiation, nuclear interactions and pair production create
charged particles each of which contribute to the Cerenkov radiation. Energy reconstruction in this regime will be
logarithmic in energy with an error of $\log_{10} \sigma_{E} \sim
0.3$~\cite{Ereconpaper}. Meanwhile, below $\sim 300 \; \mathrm{GeV}$ the length of the track is proportional to the energy of the incident
particle since the majority of energy loss in this range is governed by ionization. In this case reconstruction will likely
be more accurate for contained tracks, and reconstruction algorithms are currently in development~\cite{IceCube:2011ah}. Events between
these regimes call for a more complicated reconstruction algorithm.

Whereas the IC design is focused on event energies above a TeV, IC/DC achieves an energy threshold as low as 10
GeV. IC/DC consists of eight more densely instrumented strings with high quantum-efficiency digital optical modules (DOMs). Surrounding
IC/DC strings veto through-going muons so IC/DC records muon-neutrino vertex events inside its volume. IC/DC
increases the IC effective volume at energies below 65 GeV and accounts for the majority of events recorded
below 100 GeV. Further infills of the IC/DC array, such as PINGU, could extend the energy threshold to a few GeV and further increase the
effective volume by a factor of two at 10 GeV~\cite{DeYoung:2011ke}.

%%%%%%%%%%%%%%%%%%%%

\section{Primary Neutrinos from Dark Matter Annihilation: Theoretical Motivation}

DM particles can in principle annihilate into any of the SM particles. The annihilation rate to a final state is given by
$\sigma_{\rm ann} v = a + b v^2$, where $v$ is the relative velocity of annihilating particles. The two terms in this expression represent
the $S$-wave and $P$-wave contributions respectively. For DM annihilation inside the sun the thermalized velocity is very low ($v
\sim 10^{-4}$ for a 100 GeV DM particle). Therefore channels that proceed through the $S$-wave dominate annihilation (unless $a$ is
extremely small).

For the popular and extensively studied neutralino DM in the MSSM, annihilation is mainly into gauge boson and charged
fermion final states. For example, in the stau-neutralino coannihilation region of minimal supergravity (mSUGRA) models~\cite{coann}, where
DM is mostly Bino, annihilation in the $S$-wave of DM particles is typically dominated by the tau final states. Taus produce
relatively soft neutrinos via three-body decay. In the focus point region of mSUGRA models~\cite{focus}, where DM particles have a large
Higgsino fraction, $S$-wave annihilation is predominantly into $W$ bosons (and $t$-quarks if kinematically allowed). $W$'s produce neutrinos
with a harder spectrum via two-body decay. As a result, annihilation of neutralino DM typically yields secondary neutrinos from $W$ and
tau final states. The neutrino signal that is produced from neutralino annihilaiton inside the sun has been studied in various cases~\cite{MSSM}. Because of the existence of charged particles in the final state, annihilation of neutralino DM also results in a gamma-ray signal. This implies that
one can also obtain constraints by using data from the Fermi gamma-ray space telescope~\cite{Fermi}.

Neutralino annihilation to primary neutrinos happens via gauge interactions where neutrinos are produced through the $Z \nu {\bar \nu}$
vertex. It therefore produces a left-handed (LH) neutrino and a right-handed (RH) antineutrino. Considering that the neutralino is a
Majorana fermion, the $S$-wave contribution to such a final state is extremely small due to the tiny masses of neutrinos. We note that the
$P$-wave contribution to neutrino final states is also very small because of velocity suppression. In consequence, production of primary
neutrinos with a hard spectrum from neutralino annihilation is highly suppressed.

However, it is possible to enhance DM annihilation into primary neutrinos by going beyond neutralino DM. If DM annihilation produces a
$\nu \nu$ pair, instead of a $\nu {\bar \nu}$ pair, it can proceed in the $S$-wave without any mass suppression. A detailed analysis of
settings in which DM annihilation into neutrinos is enhanced has been given in~\cite{Lindner}. This can happen when DM is related to the
neutrino sector (for some specific models, see~\cite{Farzan1}).

One particular model that can produce hard monochromatic neutrinos is the $B-L$ extension of the MSSM ($B$ and $L$ are baryon and lepton
number respectively). A minimal extension of the SM gauge group, motivated by the nonzero neutrino masses, includes a gauged $U(1)_{B-L}$
gauge symmetry~\cite{mohapatra}. Anomaly cancellation then implies the existence of three RH neutrinos (N), the lightest superpartner of
which ($\widetilde{N}$) makes a viable DM candidate~\cite{ADRS2}.
Thermal freeze-out of the sneutrinos can yield the correct dark matter abundance if the $B-L$ symmetry is broken around the TeV
scale~\cite{ADM}. The sneutrinos can annihilate in the $S$-wave to RH neutrinos, which in turn decay to LH neutrinos and the SM Higgs. The
energy of the resulting neutrinos will be close to monochromatic as long as the difference between the masses of the RH sneutrino and RH
neutrino is much less than the RH sneutrino mass. The prospects for DM detection for this model using IC/DC is considered
in~\cite{ABDR}.\footnote{This model has other interesting cosmological and phenomenological implications.
The Tevatron~\cite{Tevatron} and LEP~\cite{LEP} limits on the $Z^{\prime}$ mass bound the RH sneutrino-nucleon scattering
cross section to be $\sigma < 7 \times 10^{-9}$ pb, which is just below the bound from the XENON experiment~\cite{Aprile:2011hi}.
Significant
DM annihilation to taus is also possible~\cite{ADRS2} (also see~\cite{ADRS1}), which could provide an explanation for the positron excess in
the cosmic rays reported by PAMELA~\cite{PAMELA}. In addition, this model can accommodate inflation~\cite{AKM}, provide a unified picture of
inflation and dark matter~\cite{ADM}, and give rise to interesting predictions for neutrinoless double beta decay experiments~\cite{rabi}.}

The neutrino signal is the main channel of indirect searches for models with DM annihilation to primary neutrinos. These models result in a
highly suppressed gamma-ray signal, which escapes the bounds set by Fermi~\cite{Fermi}. It is worth pointing out that neutrinos from DM
annihilation in the
galactic center provide complementary information to those from annihilation inside the sun: the former constrains the DM
annihilation cross section~\cite{IceCube:2011ae}, while the latter bounds the DM scattering cross section off
nucleons~\cite{Collaboration:2011ec}. The signature of DM annihilation to neutrinos at the galactic center should be distinctive
with a hard cutoff (for example, see~\cite{Erkoca:2010qx}).

%%%%%%%%%%%%%%%%%%%%
\section{Distinguishing Neutrino Final State from $W$ and Tau Final States}

MSSM neutralino annihilation predominantly produces gauge bosons and charged fermions, each of which in turn produce
secondary neutrinos via two-body or three-body decays. In this work, we adopt a model-independent approach and consider interesting
neutrino spectra from DM annihilation to prompt $\nu$'s, $W$'s, and $\tau$'s.

For charged fermions, we investigate only annihilation to $\tau$'s for the following reasons. $e$'s and $\mu$'s are stopped
immediately in the sun, so $\mu$ decay will occur at low energy and result in a very soft spectrum. When DM annihilation
produces quarks, all quarks except the $t$-quark hadronize before their subsequent decay. The $t$-quark decays to a $W$ boson and a
$b$-quark before hadronization, and the subsequent $W$ decay produces a neutrino spectrum comparable to that from the $W$ final state. For
the remaining hadronizing quarks, the lighter the quark the longer it will take to decay, and hence the lower its energy will be. The
$b$-quark produces neutrinos via three-body decay, which has a softer spectrum than that from $\tau$. All other quarks decay effectively at
rest and have unmeasurable spectra at low energies that cannot compete with the atmospheric background~\cite{Jungman:1995df}.

With regard to gauge boson final states, $W$ and $Z$ spectra are relatively similar; we consider only $W$ final states
in our analysis. Finally, the Higgs boson final states also result in a soft neutrino spectrum that is negligible compared to that from $W$
and $\tau$ final states. The Higgs mainly decays to $b$-quarks, thus yielding a neutrino spectrum similar to that from the $b$-quark final
state.

In the following, we therefore focus on secondary neutrinos from DM annihilation to $W$ and $\tau$ final states vs primary neutrinos and
their corresponding spectra. Previous investigations have shown that measuring the spectrum can allow the reconstruction of the DM mass and
its annihilation branching ratios~\cite{Cirelli} (also see~\cite{Mena} for reconstructing DM properties, and~\cite{Vernon} for the signature of primary neutrinos). We perform an analysis for reconstructing the DM elastic scattering cross
section and its annihilation channels, including experimental error in energy reconstruction and assuming that its mass can be determined
from other experiments (notably by the LHC).

%%%%%%%%%%%%%%%%%%%%%%%%%%%%%%%%%%%%%%%%%%%%%%%%
\subsection{Neutrino \& Contained Muon Spectra}

In this analysis, we consider the spectrum of contained muon tracks so that the vertex at which $\nu_\mu$ is converted
to a $\mu$ is within the detector volume. Hadronic and electromagnetic cascade events in the ice also carry useful information about low
energy neutrinos and may show tau neutrino appearance from oscillations~\cite{tauAppearance}. IC/DC is uniquely suited to measure
these low energy events and recently confirmed the observation of neutrino-induced cascades~\cite{DeYoung:2011ke}. For our purposes,
since cascades are localized in the ice and thus suffer from an angular resolution of about $60$ degrees at $100$ GeV, too much background
is admitted to perform a meaningful DM search with cascades~\cite{cascadeAngle}.\footnote{Muon tracks and cascades together can be used to gain information about the flavor ratio of neutrinos at the detector, which can be used as diagnostic of DM annihilation inside the sun~\cite{Weiler}.}

We use DarkSUSY to calculate the spectrum for any given final state~\cite{DarkSUSY}. To the extent that the energy of the event may be well
reconstructed in this range, we also assume that the muon track itself is fully contained, ending inside the
detector. We do not account for any loss of events or poor energy reconstruction at the edges of the fiducial volume.

\begin{figure}[h!]
% \ContinuedFloat
  \centering
  \subfloat[{\bf {Annihilation to $\nu$ final state}}]{\label{fig:1a}\includegraphics[width=.42\textwidth]{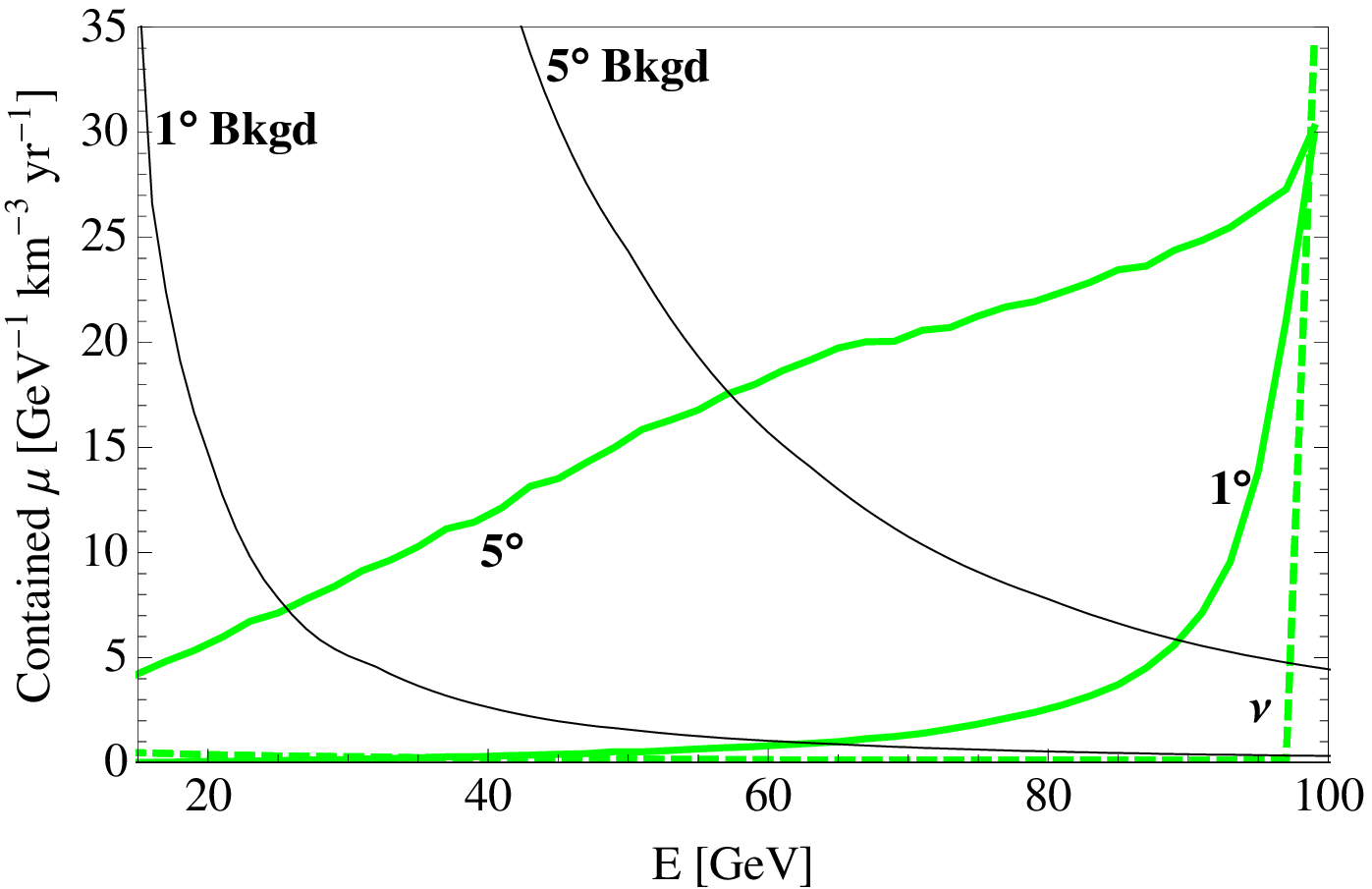}} \,
\linebreak
  \subfloat[{\bf {Annihilation to $W$ final state}}]{\label{fig:1b}\includegraphics[width=.42\textwidth]{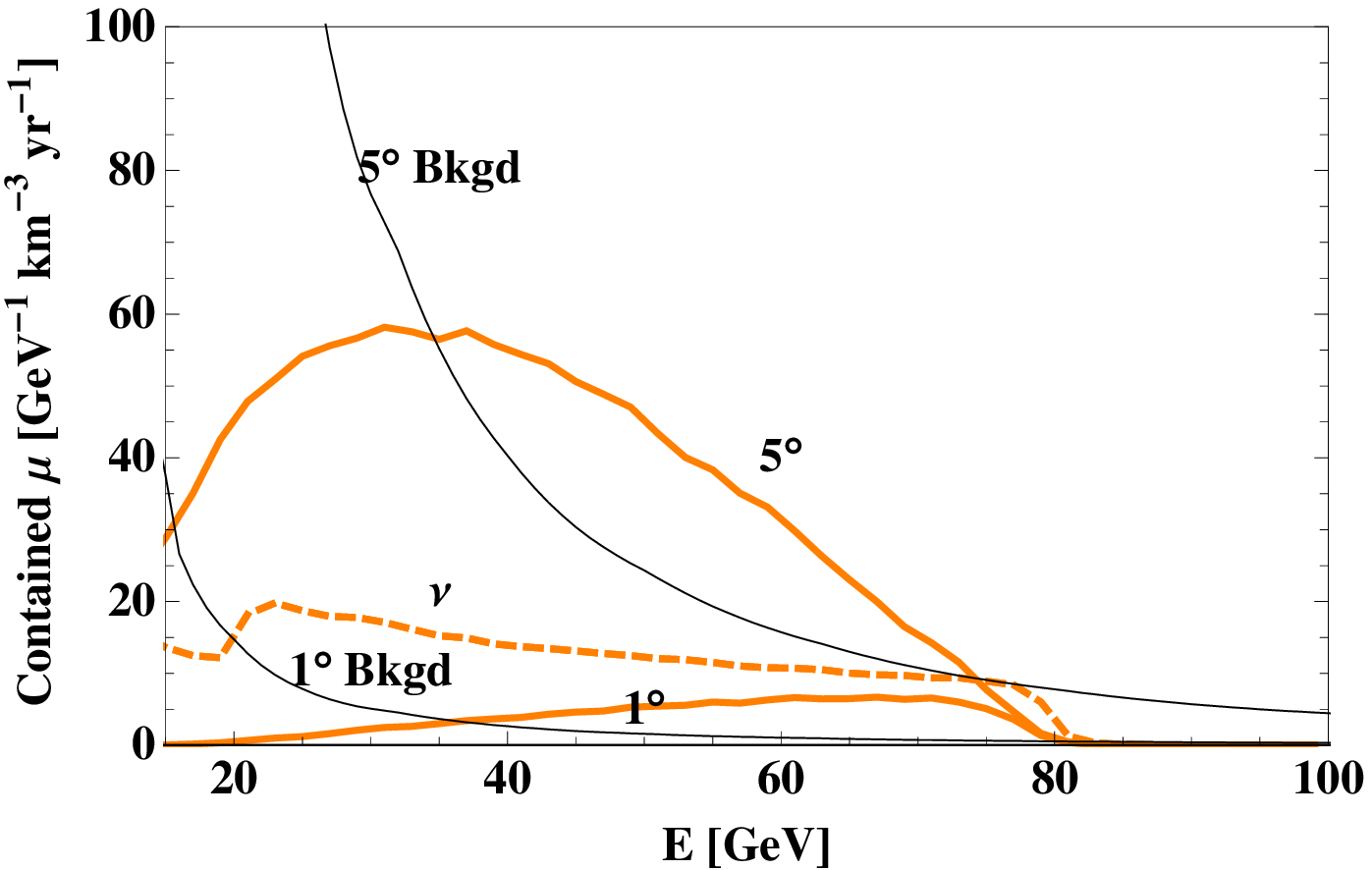}} \,
\linebreak
\vspace{.5ex}
  \subfloat[{\bf {Annihilation to $\tau$ final state}}]{\label{fig:1c}\includegraphics[width=.42\textwidth]{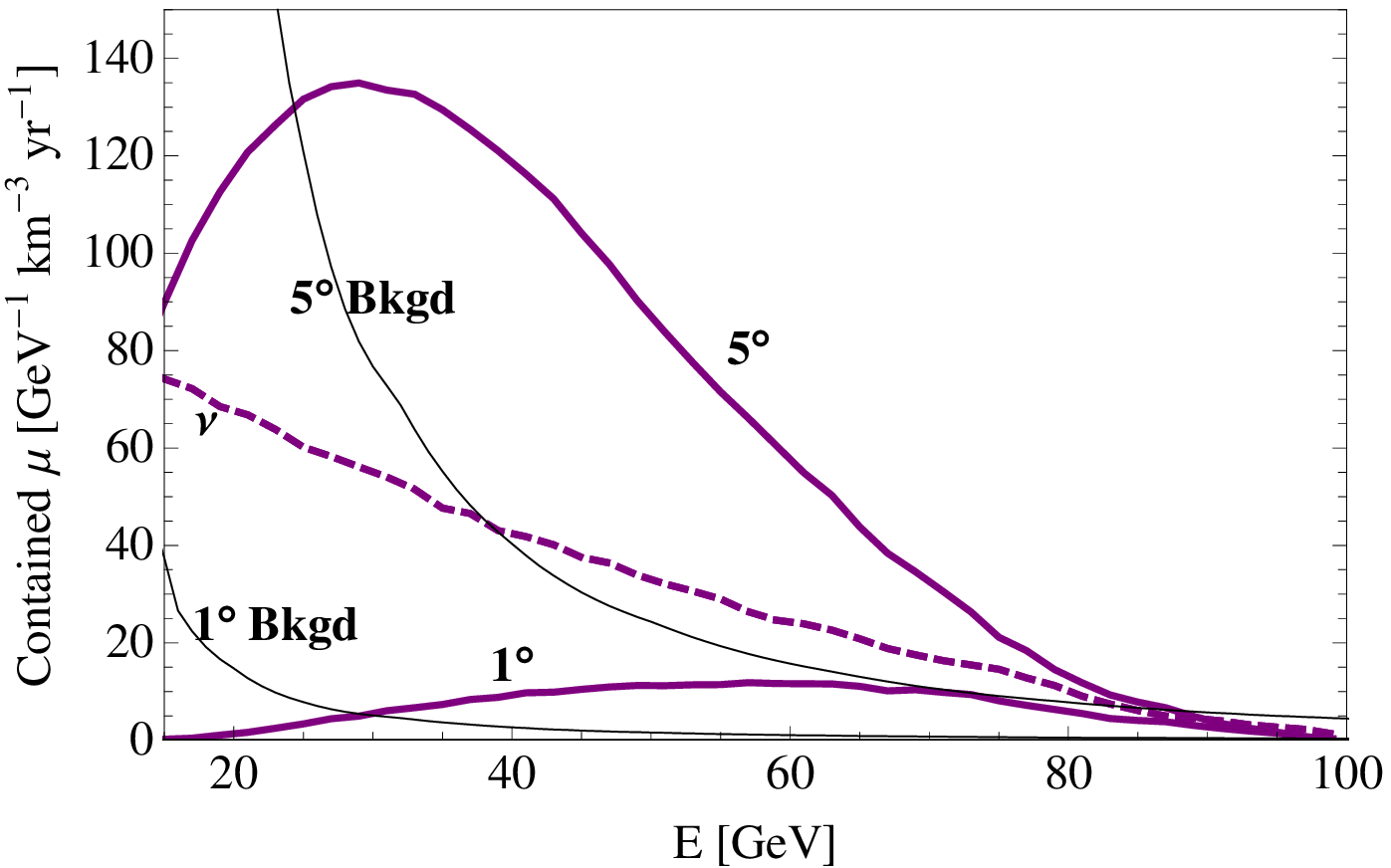}} \,
\caption[Neutrino \& Muon Spectra]{Energy spectra of contained muon events (solid) and incident neutrinos (dashed) at the earth from a
$100$ GeV DM particle for various annihilation channels. Relative to the sun $1^{\circ},~5^\circ$
angular cuts have been placed on the signal and atmospheric background muons (thin black lines). The top plot for annihilation
to neutrinos is for $\sigma_{\rm SD}= 10^{-41} \; \mathrm{cm}^2$, while the other plots are for $\sigma_{\rm SD}=
10^{-40} \; \mathrm{cm}^2$, all of which are below the current experimental bounds. Normal mass hierarchy and $\theta_{13} = 10^\circ$ have
been chosen here, but the spectra are largely insensitive to the oscillation parameters.}
\label{fig:1}

\end{figure}

As an example, Fig.~\ref{fig:1} depicts the theoretical prediction for neutrino and muon spectra that result from the annihilation of
$100$ GeV DM particles in the sun to prompt $\nu_\mu$'s, $W$'s and $\tau$'s. The flux is given in events per km$^{3}$ effective volume of
the
detector
at earth per year. Only a half-year of data is taken for each calendar year, since we assume the neutrino telescope requires the sun to
be below the horizon to observe the DM neutrino signal above the atmospheric muon background. For a DM mass of $100$ GeV the muons
will be fully contained and accessible in IC. The qualitative features of the neutrino and muon spectra (such as peaks and kinematic
cutoffs) are the same for higher DM masses.

The prompt neutrinos, Fig.~\ref{fig:1a}, result in a monochromatic peak at the DM mass in the neutrino spectrum. The
contained muon spectrum is plotted along with the neutrino spectrum to emphasize the effect of the charged current conversion of neutrinos
to muons. The conversion demonstrates the weak scale reduction in events relative to the incident neutrinos and is proportional to energy,
so the peak from the neutrino annihilation channel is well preserved. The spectra are largely insensitive to the choice of
neutrino oscillation parameters and annihilating neutrino flavor. Here we depict $\theta_{13} = 10^\circ$, normal mass hierarchy and
annihilation to $\nu_\mu$, but scenarios with inverted mass hierarchy, $0 \leq \theta_{13} < 10^\circ$ and other neutrino flavors in the
final state only slightly change the height of the peak.\footnote{Recent results from T2K~\cite{T2K} and MINOS~\cite{MINOS} experiments
suggest nonzero $\theta_{13}$ at $2.5\sigma$ and $89\%$ confidence respectively. The
results are consistent with $\theta_{13} = 10^\circ$.}

The $W$ bosons produce neutrinos via two-body decay, which are softer than the previous case. For a highly boosted $W$ (whose energy is
equal to the DM mass), the energy of secondary neutrinos {\bf is between $M_{DM}(1 - \beta)/2$ and $M_{DM}(1 + \beta)/2$, where $\beta = \sqrt{1 - (M_{W}/M_{DM})^2}$}. This is the reason for the
relatively sharp kinematic edges in the neutrino spectrum in Fig.~\ref{fig:1b}. We note that $W$'s also decay to $b$-quarks, which in turn
produce tertiary neutrinos. This results in additional contributions to the neutrino spectrum {below the lower
kinematic cutoff}. The muon spectrum has a peak at energies well below the DM mass resulting from charged current
interactions whose cross section is proportional to the neutrino energy.

{An important point to note is that $W$'s with transverse polarization can dominate the annihilation final state as happens, for
example, in the case of neutralino DM. The main contribution to hard neutrinos (with an energy close to the upper kinematic cut off
mentioned above) comes from transverse $W$'s in this case, which will become totally dominant when $M_{DM} \gg M_W$. In these cases, it is
therefore important to carry out a spin-dependent calculation that accounts for spin correlations of the final state particles and the
helicity-dependence of their decays~\cite{Danny}. We also note that the upper kinematic cut off itself gets closer to $M_{DM}$, and thus
creates a harder neutrino spectrum, when $M_{DM} \gg M_W$.}

The $\tau$'s produce neutrinos via three-body decays, which results in a much softer neutrino spectrum that rises toward lower energies as
seen in Fig.~\ref{fig:1c}. Thus the peak of the resulting muon spectrum is located at a lower energy than that in Fig.~\ref{fig:1b}.
However, there is no kinematic cutoff in the neutrino spectrum in this case, and hence the muon spectrum extends smoothly to the DM mass.

The important point is that there is a distinct separation in energy between the contained muon peak of prompt neutrinos at the DM mass and
the $W$ and $\tau$ peaks at lower energies. This indicates that scenarios with DM annihilation to primary neutrinos can be
distinguished from those with DM annihilating to $W$ bosons or $\tau$'s. Moreover, were a model to contain both a
neutrino final state and a $W$ or $\tau$ final state, the neutrino signal may be used to determine the corresponding branching ratios.

%%%%%%%%%%%%%%%%%%%%%%%%%%%%%%%%%%%%%%%%%%%%%%%%%%%%%%%%%
\subsection{Reconstruction of Annihilation Channels}

\begin{figure}[h!]
\centering
\includegraphics[width=.47\textwidth]{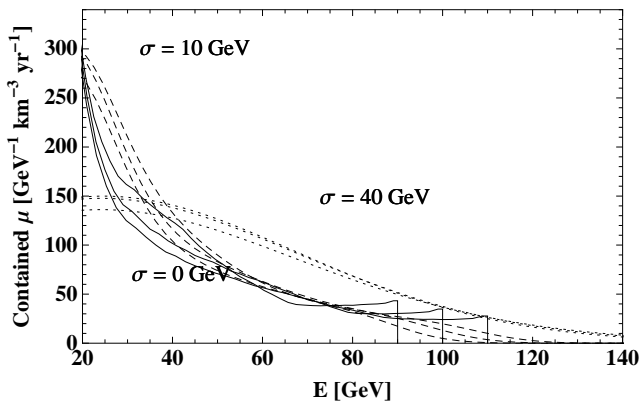}
\caption[Added Spectra]{\label{fig:2} Total contained muon spectrum from 90, 100 and 110 GeV DM particles (from left to right) with $\sigma_{\rm SD} = 10^{-40}$
cm$^2$ annihilating to $W$'s ($90\%$) and $\nu$'s ($10\%$) with the atmospheric background added. A $5^\circ$ cut has been
placed on the events. Smeared spectra for energy reconstruction errors of 10 GeV (dashed) and 40 GeV (dotted) are also shown.}
\end{figure}

The theoretical predictions shown in the previous subsection are subject to experimental error. Individual muon events from a
particular annihilation channel cannot be tagged as such, nor can individual signal and background
events be distinguished. A more realistic picture is shown in Fig.~\ref{fig:2} for a 100 GeV DM particle, where two channels along
with background events have been added together and subjected to a $5^\circ$ angular cut. The branching ratios are $90\%$ to $W$ bosons
and $10 \%$ to neutrinos with $\sigma_{\rm SD} = 10^{-40} \mathrm{cm}^2$. This is a factor of a few below the current bounds from
IC~\cite{Collaboration:2011ec} and just at the current bounds of Super Kamiokande (SuperK) for annihilation to
$W$'s~\cite{Tanaka:2011uf}.\footnote{For a purely spin-independent cross section, the same result is obtained for $\sigma_{\rm SI} \approx 3
\times 10^{-43}$ cm$^2$. However, such a large value of $\sigma_{\rm SI}$ is already ruled out by direct detection
experiments~\cite{Aprile:2011hi}. For this reason we focus on the spin-dependent cross sections in this analysis.}

Angular cuts on the muon tracks relative to the position of the sun lower the relevant atmospheric contained muon background relative to
the signal. Lower energy events are preferentially scattered at higher angles relative to the incoming neutrino; therefore, angular cuts
disproportionately remove low energy events. A $1^\circ$ cut leaves the majority of the high energy muons from primary neutrinos, which are
clearly separated from a reduced background. Meanwhile, a $5^\circ$ cut admits more lower energy muons that are indicative of the
pronounced peaks from $W$ and $\tau$ final states. Different angular cuts will optimize the
signal to background ratio for different annihilation channels. We have found a $5^\circ$ cut to be optimal, where the background does not
overwhelm the signal in the regime above 40 GeV and a significant portion of the $W$ peak remains. The contribution from neutrinos is still
easily distinguished in the peak cutoff.

The realistic energy reconstruction for a contained muon spectrum at IC/DC for these energies is not certain. Energy reconstruction for
through-going TeV muons is approximately $\sigma_{\log_{10} E} = 0.3$ \cite{Ereconpaper}. Energy loss of the muon is described by
\begin{equation}
\frac{dE}{dx} = a + b E \;,
\end{equation}
where $a$ quantifies muon loss via ionization, and $b$ quantifies loss from pair productions, Brehmsstrahlung radiation and nuclear
interactions. For ice at energies above 1 TeV, enough Brehmsstrahlung and pair production occurs to ensure that the light deposited
in the detector is proportional to the muon energy. However, for contained muon events less than one kilometer in extent,
the track length is proportional to the muon energy. Reconstruction in this regime will be better than in the regime above 1 TeV
but will depend on the geometry of the detector.

The energy resolution error of the IC/DC detector for contained muon events {about} 100 GeV should be linear with energy, in that the
track
length is proportional to the energy of the incoming muon. In Fig.~\ref{fig:2}, we recreate a Gaussian energy reconstruction with an error,
or width, equal to $10$ or $40$ GeV. {This is equivalent to claiming that the track length can be known to within $50-200$ m, a resolution
that seems feasible for IC/DC (or telescopes withy similar capabilities) since IC/DC DOMs are spaced at $7$ m vertically and strings are $72$ m apart.}
%while IC DOMs are spaced at $17$ m vertically and strings are $125$ m apart. {\bf Should we just say IC/DC here?}

\begin{figure}[h!]
% \ContinuedFloat
  \centering
  \subfloat[{\bf {10 GeV Energy Resolution Error}}]{\label{fig:3a}\includegraphics[width=.42\textwidth]{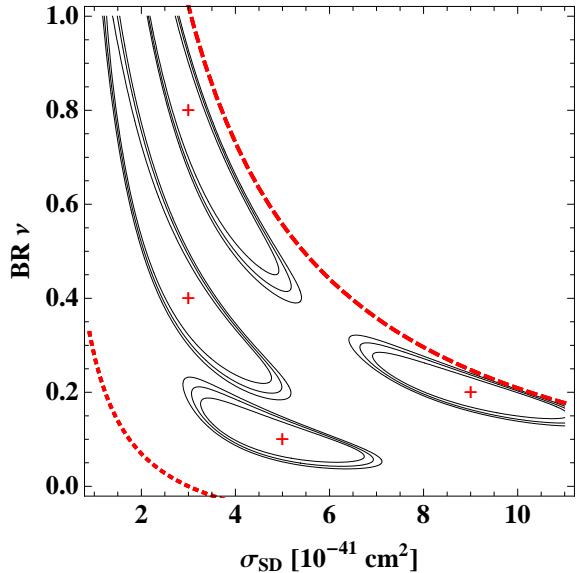}} \,
\linebreak
\vspace{.5ex}
  \subfloat[{\bf {40 GeV Energy Resolution Error}}]{\label{fig:3b}\includegraphics[width=.42\textwidth]{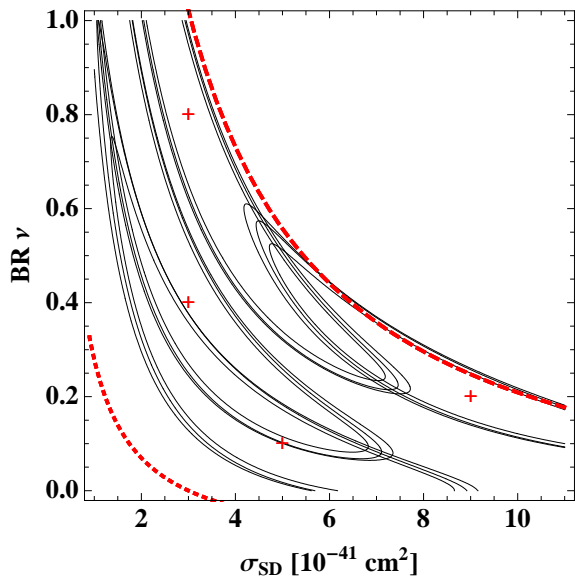}} \,
\caption[Reconstruction Confidence]{\label{fig:3} Contours of $2 \sigma$ confidence for reconstructing spin-dependent cross section
$\sigma_{\rm SD}$ and branching ratio to prompt neutrinos ${\rm BR}_\nu$ for DM particle annihilating to $\nu$'s and
$W$'s. Contours for 90 (inner), 100 (middle) and 110 GeV (outer) DM are shown. One
year of data with km$^3$ effective volume is used with energy reconstruction errors of 10 GeV (top) or 40 GeV (bottom). The region of interest is between the current
SuperK upper bound (dashed)~\cite{Tanaka:2011uf} and the future IC86 sensitivity bound (dotted)~\cite{IceCube:2011ae}.}
\end{figure}

{The experimental error in measuring the DM mass at colliders should also be taken into account. It has been shown that in some
supersymmetric scenarios a $10\%$ accuracy can be reached at the LHC~\cite{Bhaskar}. For a 100 GeV DM particle this corresponds to an error
of 10 GeV, and hence we include 90, 100 and 110 GeV masses in our analysis.}

It is seen that the smoothing of the spectrum considerably suppresses the distinct
channel features, especially for a 40 GeV error in energy reconstruction. Nevertheless, it is still possible to reconstruct the total
spin-dependent scattering cross section $\sigma_{\rm SD}$ and the branching ratio for annihilation to primary neutrinos ${\rm BR}_\nu$. 
{We also note that the uncertainly in determination of the DM mass is subdominant to the experimental error in energy reconstruction, which
results in minimal changes in the smoothed spectra.}

In Fig.~\ref{fig:3} we show the $2 \sigma$ confidence contours for 90, 100 and 110 GeV DM particles annihilating to $\nu$ and $W$ final
states, using one year of data and applying energy reconstruction errors of 10 GeV and 40 GeV. These contours account for Poisson errors,
the addition of background and a 40 GeV energy cut; we have used the $\chi^2$ analysis in~\cite{Friedland1999} to obtain these results. {We
again see that varying DM mass from 90 GeV to 110 GeV produces minimal changes in the contours.}

The plot in Fig.~\ref{fig:3} also shows the current SuperK bound~\cite{Tanaka:2011uf} and the future IC/DC sensitivity
bound~\cite{IceCube:2011ae} on $\sigma_{\rm SD}$ (dashed and dotted curves respectively). For $100\%$ DM annihilation to $W$'s, the
experiments have derived bounds on $\sigma_{\rm SD}$ by applying appropriate energy and angular cuts on the muon spectrum. No limit has been
placed yet for DM annihilation to primary neutrinos by either experiment. In the absence of a thorough analysis, we have used a simple
criterion to find an approximate bound on $\sigma_{\rm SD}$ for the neutrino final states. We compared the total number of contained muon
events with a $40$ GeV energy threshold and $5^\circ$ angular cut for the two final states. In the case of primary neutrinos, the flux of
$\mu$'s is about eight times larger than the flux from $W$'s, as expected from the harder neutrino spectrum. We assumed the limit on
$\sigma_{\rm SD}$ in this case is also tighter by the same factor. A dedicated analysis for the neutrino final state, with optimized angular
and energy cuts, would result in a more precise bound. Indeed, theoretical motivation for models with DM annihilation to primary neutrinos
warrants such a study by experimental collaborations. In the presence of both $\nu$ and $W$ final states, the limits are approximately given
by
\begin{eqnarray}
\sigma_{\rm SD} \left(1 + 7 ~ {\rm BR}_\nu \right)  \leq \sigma_{\rm max} \, .
\end{eqnarray}
The current bound from SuperK~\cite{Tanaka:2011uf} and the future IC/DC sensitivity bound~\cite{IceCube:2011ae} correspond to $\sigma_{\rm
max}
\approx 2.5 \times 10^{-40}$ and $4 \times 10^{-41}$ cm$^2$ respectively.

It is seen from the shape of the contours that higher $\sigma_{\rm SD}$ and smaller ${\rm BR}_\nu$ are difficult to distinguish from lower
$\sigma_{\rm SD}$ and larger ${\rm BR}_\nu$. The number of muon events at the DM mass mainly comes from the peak of primary neutrinos, which
is determined by ${\rm BR}_\nu$. This holds after smoothing of the spectrum because of the kinematic cutoff that appears well
below the DM mass for annihilation to $W$'s, see Fig.~\ref{fig:1b}. This implies that after smearing the main contribution to the total
spectrum at energies around and above the DM mass still comes from primary neutrinos. On the other hand, the main contribution to lower
energy muons comes from secondary neutrinos produced by $W$ decay. For small values of $\sigma_{\rm SD}$ or large values of ${\rm BR}_\nu$,
this contribution is small and overwhelmed by the Poisson error from the atmospheric background. Therefore, one can compensate for a change
in $\sigma_{\rm SD}$ by a corresponding change in ${\rm BR}_\nu$ and obtain spectra that are statistically indistinguishable. For larger
values of $\sigma_{\rm SD}$ or smaller values of ${\rm BR}_\nu$, the contribution of secondary neutrinos becomes significant, which limits
simultaneous variations in $\sigma_{\rm SD}$ and ${\rm BR}_\nu$ that keep the peak height unchanged. As a result, the $2 \sigma$ contours
are tighter for reconstruction points toward the right and bottom of the plot in Fig.~\ref{fig:3}. As expected, an energy
reconstruction error of 40 GeV does a poorer job because it further suppresses the features discussed above.

\begin{figure}[h!]
% \ContinuedFloat
  \centering
  \subfloat[{\bf {10 GeV Energy Resolution Error}}]{\label{fig:3a}\includegraphics[width=.42\textwidth]{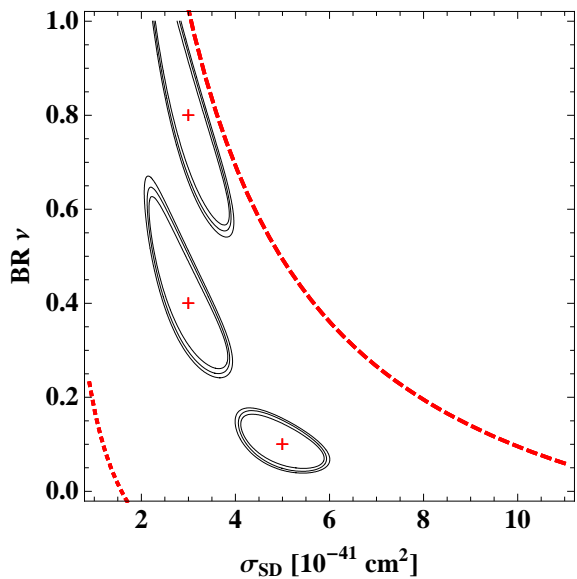}} \,
\linebreak
\vspace{.5ex}
  \subfloat[{\bf {40 GeV Energy Resolution Error}}]{\label{fig:3b}\includegraphics[width=.42\textwidth]{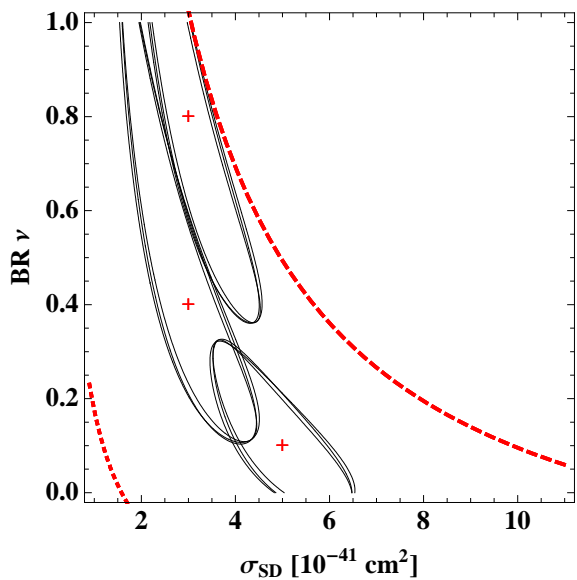}} \,
\caption[Reconstruction Confidence]{\label{fig:4} The same as Fig.~\ref{fig:3}, but for annihilation to $\nu$ and $\tau$ final
states.}\end{figure}

In Fig.~\ref{fig:4} we show the $2 \sigma$ confidence contours from a similar analysis for annihilation to $\nu$'s and $\tau$'s. {The
contours are narrower in this case, implying that the reconstruction is easier. The main reason for this is that the number of muon events
from secondary neutrinos is significantly larger, compare Figs.~\ref{fig:1b} and~\ref{fig:1c}. The larger signal to background
ratio places a stronger constraint on the simultaneous variation of ${\rm BR}_\nu$ and $\sigma_{\rm SD}$, which results in an improved
reconstruction.}
%This is due to the different shapes for
%the $W$ and $\tau$ final state spectra, see Figs.~\ref{fig:1b} and~\ref{fig:1c}. In the case of the $\tau$ final state, the spectrum
%resulting from secondary neutrinos is softer and extends all the way to the DM mass. Therefore secondary neutrinos
%can make a significant contribution to the total spectrum at energies around and above the DM mass after energy reconstruction and are also
%affected more by the background at lower energies. Both of these effects imply a poorer reconstruction of the branching ratios in this case.

Some comments are in order. In producing these figures, a scan of angular cuts in $1^\circ$ increments within the $1^\circ-10^\circ$ range
was made. While visually a $1^\circ$ cut may be optimal in distinguishing the presence of the neutrino channel, a $5^\circ$ cut was optimal
in reconstructing the branching ratios in the presence of the broad, low energy $W$ or $\tau$ spectrum.

We also note that the muon event rates for both the signal and background must be convolved with the effective volume of the detector for
contained muon events, which is a function of energy. IC/DC maintains significant volume above 10 GeV~\cite{DeYoung:2011ke}. While the
geometric volume for IC/DC is about $3\%$ of the volume of IC, the more dense spacing of DOMs in IC/DC make it more efficient at event
detections for energies below 100 GeV~\cite{DeYoung:2011ke}. The effective volume in this range increases with energy as longer muon tracks
are more likely to produce detectable light. This can make the effect of the peak at the DM mass from primary neutrinos even more
distinctive above the smeared $W$ or $\tau$ spectrum. The reconstruction can therefore be better than shown here after accounting for the
energy dependence of the effective volume.

Finally, one year results for a $\mathrm{km}^3$ effective volume, with the sun below the horizon, translate to roughly ten years of a
$0.05 \; \mathrm{km}^3$ detector capable of operating for a full year. Thus, in the region of IC/DC sensitivity that would yield a
discovery of prompt neutrinos after one year, 10 years of data could allow reconstruction of the branching ratios for
the IC/DC effective volume. A larger detector with the same capabilities can significantly improve these prospects.

%%%%%%%%%%%%%%%%%%%%%%%%%%%%%%%%%%%%%%%%%%%%%%%%%%%%%%%%%%%%%%%%%%%%%%%%%
\section{Distinguishing Neutrino Flavors} \label{ChannelContributions}

Once the presence of direct DM annihilation to neutrinos is confirmed, it is desirable to also learn
the flavor of the final state neutrinos. This will provide specific information about models that connect DM to the neutrino sector.
For example, in the $U(1)_{B-L}$ extension of MSSM where the lightest RH sneutrino is the DM candidate~\cite{ADRS2}, flavor composition of
final state neutrinos is related to the neutrino Yukawa couplings. Therefore, knowledge of the neutrino flavors will yield useful
information pertaining to the underlying neutrino mass model and leptogenesis.

Here we show how regeneration of $\nu_\tau$ inside the sun and its effect on the muon spectrum may be used as a novel method to distinguish
the flavor of primary neutrinos in the final state.\footnote{It has been proposed that seasonal variation of the neutrino signal may also be
used to extract information about the flavor of primary neutrinos~\cite{EF}.}

%%%%%%%%%%%%%%%%%%%%
\subsection{The $\nu_\tau$ Regeneration Effect}
DM annihilation produces neutrinos in flavor eigenstates in the sun. Neutrinos then undergo charged current interactions with matter
as they propagate through the sun. These interactions convert $\nu_e,~\nu_\mu,~\nu_\tau$ to $e,~\mu,~\tau$ respectively. $e$'s and $\mu$'s
are stopped immediately due to electromagnetic interactions. On the other hand, the $\tau$ decays quickly before losing too much energy
because of its very short lifetime of $3 \times 10^{-13} \; \mathrm{s}$~\cite{pdg}. This decay produces a $\nu_\tau$, which has a
lower energy than the original one. Charged current interactions therefore suppress the peak of the neutrino spectrum at the DM mass for all
flavors.\footnote{Neutrinos also have neutral current interactions with matter inside the sun. Scatterings via neutral current interactions
result in energy loss of the neutrinos and further suppress the peak at the DM mass. However, the cross section for neutral current
interactions is a factor of 3 smaller than that for charged current interactions, which makes them subdominant. More importantly, neutral
current scatterings affect all flavors equally.} However, in the case of the $\nu_\tau$, the regeneration of neutrinos via three-body decay
populates the spectrum at energies well below the DM mass.

\begin{figure}[h!]
\centering
% \ContinuedFloat
\subfloat[{\bf {400 GeV DM mass}}]{\label{fig:5a}
\includegraphics[width=.47\textwidth]{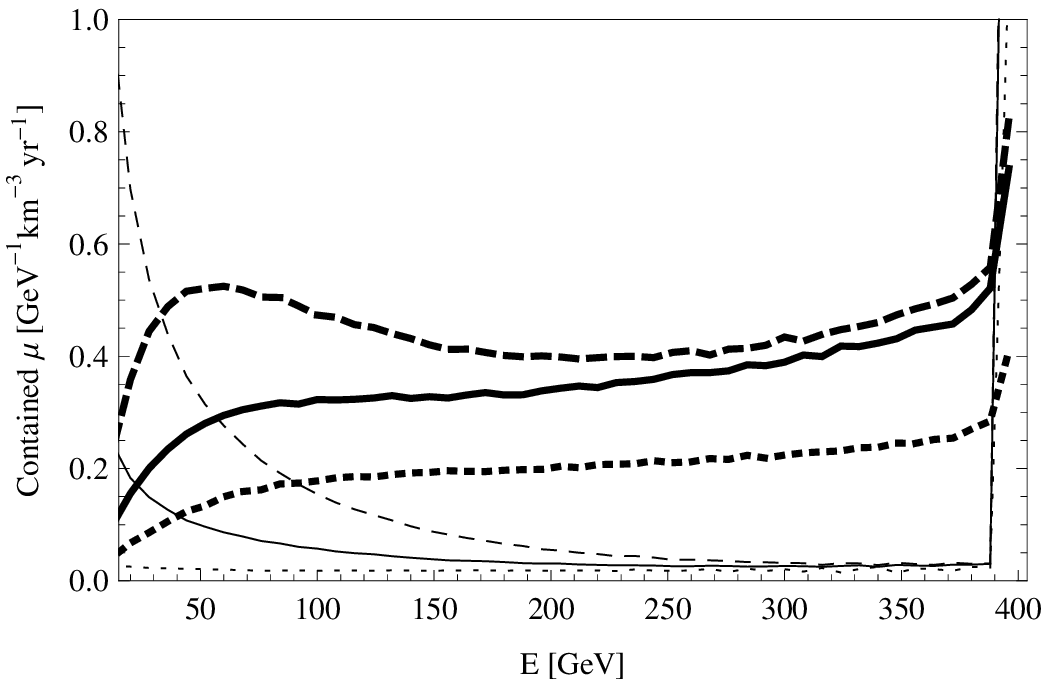}}\,\,
\linebreak
\subfloat[{\bf {600 GeV DM mass}}]{\label{fig:5b}
\includegraphics[width=.47\textwidth]{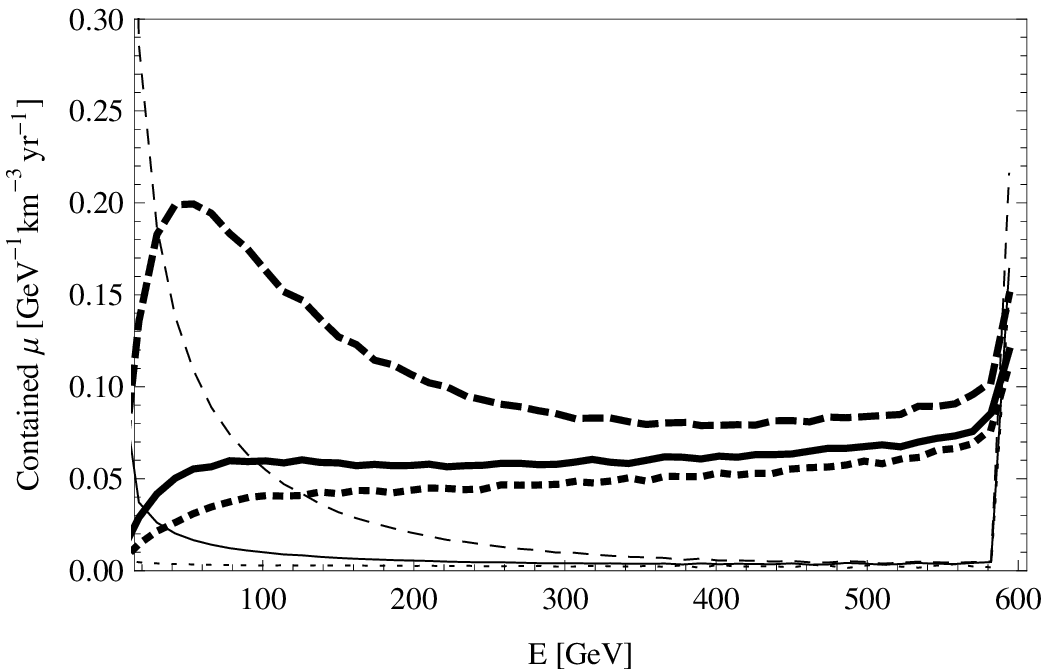}}\,\,
\caption[Tau Regeneration Spectra]{\label{fig:5} Contained muon spectra for DM annihilation to $\nu_e$ (dotted), $\nu_\mu$ (solid) and
$\nu_\tau$ (dashed) for 400 (Fig.~\ref{fig:5a}) and 600 GeV (Fig.~\ref{fig:5b}) DM masses and $\sigma_{\rm SD} = 10^{-41}$ cm$^2$. Normal
mass hierarchy and $\theta_{13}=10^\circ$ are chosen here, but substantive features of the spectra do not depend on the neutrino oscillation
scenario. A $5^\circ$ cut has been placed on the events. Corresponding neutrino spectra are shown in thin lines.
The regeneration effect is evident for the $\nu_\tau$ final state and results in a peak in the muon spectrum at low energies. For a 400 GeV
DM particle, regeneration also affects the $\nu_\mu$ final state due to efficiency of $\nu_\mu-\nu_\tau$ oscillations inside the sun.
}
\end{figure}

The cross section for charged current interactions is proportional to the energy of primary neutrinos produced at the center of the sun,
which is essentially equal to the DM mass. Neutrino absorption becomes significant when the absorption length of neutrinos $L_{\rm abs}$ is
roughly equivalent to the core size of the sun $R_C \sim 70,000$ km. Using the charged current neutrino-nucleon cross
section~\cite{CooperSarkar:2007cv} and a core density of $150 \; \mathrm{g/cm}^3$, we find that $L_{\rm abs} \sim 70,000$ km at energies
$E_\nu \sim 300$ GeV. Oscillations among different flavors should be taken into account as neutrinos travel through the sun.
We note that for $\nu_e$ the flavor and mass eigenstates are the same deep inside the sun. This implies that upon production at the center
of the sun $\nu_e$'s propagate through the core without changing their flavor. Thus absorption via charged current interactions start to
suppress the flux of $\nu_e$ for DM masses above $300$ GeV as discussed above.

On the other hand, $\nu_\mu$ and $\nu_\tau$ are not mass eigenstates inside the sun, and hence undergo oscillations. Since matter effects
are the same for these two flavors, the $\nu_\mu-\nu_\tau$ oscillation length inside the sun is set by the atmospheric mass splitting
$L_{\rm osc} = E_\nu/4 \pi \Delta m^2_{\rm atm}$. As long as $L_{\rm abs} \gs L_{\rm osc}/4$, oscillations mix $\nu_\mu$ and $\nu_\tau$
efficiently before the absorption becomes important. As a result, $\nu_\mu$ final states also feel the regeneration effect. Since
$L_{\rm abs} \propto E^{-1}_\nu$ and $L_{\rm osc} \propto E_\nu$, at sufficiently high energies $L_{\rm abs}$ drops below $L_{\rm osc}/4$.
This happens for a DM mass of about $500$ GeV. Starting at this point, $\nu_\mu-\nu_\tau$ oscillations cease to be effective. Hence
$\nu_\mu$ gets absorbed through charged current interactions similar to $\nu_e$. In consequence, only the
$\nu_\tau$ final state retains a significant regeneration signature for DM masses above 500 GeV.

In Fig.~\ref{fig:5} we show the muon and neutrino spectra for different flavors of primary neutrinos. As we see in Fig.~\ref{fig:5a}, the
$\nu_\mu$ channel shows some regeneration effect for a 400 GeV  particle, which makes it distinguishable from the $\nu_e$ channel.
For a 600 GeV DM particle, see Fig.~\ref{fig:5b}, the $\nu_\mu-\nu_\tau$ oscillations are inefficient. Therefore, only the $\nu_\tau$
channel shows significant regeneration, which makes it distinguishable from the $\nu_\mu$ and $\nu_e$ channels.
The regeneration effect results in a peak in the muon spectrum at low energies, which becomes more pronounced as the DM mass increases.
These figures are for the normal hierarchy of neutrinos and $\theta_{13} = 10^\circ$, but variations in the neutrino oscillation scheme do
not change the substantive features of the spectra.

%%%%%%%%%%%%%%%%%%%%%%%%%%%%
\subsection{Signals of Neutrino Flavor}

As we saw, the $\nu_\tau$ regeneration effect becomes significant at DM masses above $300$ GeV. The background from atmospheric neutrinos is
relatively small at such energies due to the power law decrease in cosmic ray background. However, the neutrino signal from DM annihilation
is also kinematically suppressed for heavier DM masses. This leaves few events to detect in the interesting range of the spectrum where a
regeneration peak is evident. Additionally, energy reconstruction for the contained muon spectrum above $300$ GeV begins to suffer
from logarithmic error in energy since the reconstruction also depends on the amount of light produced in Brehmsstrahlung radiation and pair
production. Events cannot be fully contained at these energies, so reconstruction efforts based on track length are imprecise.\footnote{If
the energy of the corresponding cascades accompanying the muon event can also be captured, energy reconstruction would be improved.}

\begin{figure}[h!]
\centering
% \subfloat[{\bf {Integrated flux}}]{\label{fig:6a}
\includegraphics[width=.47\textwidth]{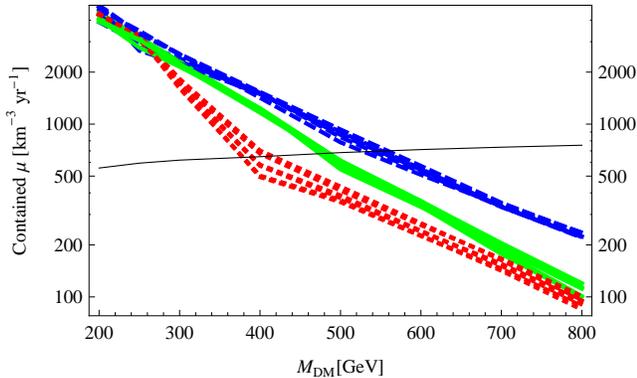}
\caption[Integrated Tau Regeneration]{\label{fig:6} Integrated muon spectra from 60 GeV to the DM mass with a $5^\circ$ angular cut as a
function of DM mass for $\sigma_{\rm SD} = 10^{-40}$ cm$^2$. The dotted, solid and dashed lines denote the $\nu_e,~\nu_\mu$
and $\nu_\tau$ final states respectively. The background is shown in black line. The bands of lines for each flavor depict different oscillation scenarios (normal and inverted
mass hierarchy as well as $0 \leq \theta_{13} \leq 10^\circ$). The separation between the $\nu_\tau$ and $\nu_e$
exceeds the Poisson error of the signal and background for DM masses above 300 GeV, and the $\nu_\tau$ becomes distinguishable
from $\nu_\mu$ above 600 GeV as well.}
\end{figure}

To mitigate these experimental challenges, we integrate the contained muon events above a threshold to consider if the cumulative effect of
the regeneration is visible. In Fig.~\ref{fig:6} we show the integrated events above $60$ GeV with a $5^\circ$ angular cut, which
readily retains the low energy regeneration effect. One can see the separation between the $\nu_e$ channel and the $\nu_\mu,~\nu_\tau$
channels above $300$ GeV, and between the $\nu_\tau$ channel and the $\nu_e,~\nu_\mu$ channels above $500$ GeV (as discussed above).
Separation between signals from different channels is typically larger than the Poisson error of the background and signal together. The
background and signal are shown separately since the background may be subtracted from the signal by observing away from the sun,
off-source, as is done in galactic center DM searches~\cite{IceCube:2011ae}. We also note that the oscillation scenario does not
significantly affect these results. The bands of lines depicting the normal and inverted hierarchies as well as $\theta_{13}$ ranging from
$0$ to $10^\circ$ for each flavor do not overlap at DM masses above $300$ GeV. These results assume $\sigma_{\rm SD} = 10^{-40}
\mathrm{cm}^2$, which is compatible with the current bounds from IC for heavier DM, and is within the reach of IC/DC sensitivity
limits~\cite{IceCube:2011ae}.

\begin{figure}[h!]
\centering
% \subfloat[{\bf {Normalized integrated flux}}]{\label{fig:6b}
\includegraphics[width=.47\textwidth]{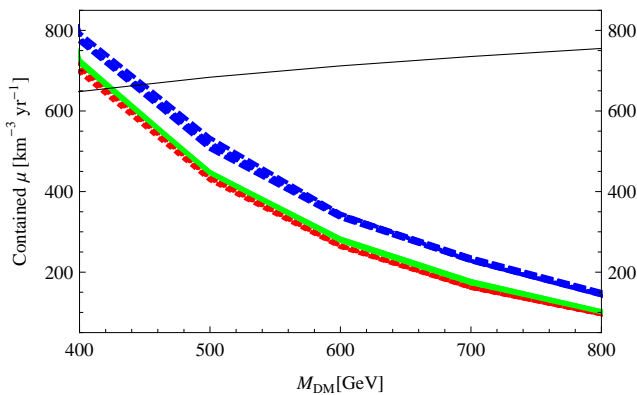}
\caption[Normalized Tau Regeneration]{\label{fig:7} The same as in Fig.~\ref{fig:6} after normalization to account for unknown value
of the DM elastic scattering cross section. The normalized spectra give the same number of events (when integrated from $200$ GeV to the DM
mass for a $1^\circ$ angular cut) as the $\nu_\mu$ final state for normal mass hierarchy and $\theta_{13}=10^\circ$. The separation between
the $\nu_\tau$ and other flavors still exceeds the Poisson error of the signal and background.}
\end{figure}

If the elastic scattering cross section of the DM is known, then a simple integration as in Fig.~\ref{fig:6} will be
sufficient to distinguish the flavor of primary neutrinos by using the regeneration effect. However, a confirmation of annihilation to
primary neutrinos would likely come from the presence of a monochromatic peak alone, the height of which depends on the scattering cross
section. In the absence of information about the cross section, we then need to normalize the signal to extract information about the
neutrino flavor. The height of the peak in the muon spectrum is closely related to the height of the neutrino peak. However, for DM masses
above 300 GeV, for which the regeneration effect becomes important, muon events at the peak are not fully contained. It will be most
difficult to reconstruct the energy of these events without track-length information. Instead, we integrate muon events above 200 GeV, which
are essentially the through-going muons, using a $1^\circ$ angular cut on the muon spectrum in order to best capture the monochromatic peak.
We then use this to normalize the integrated muon flux above 60 GeV with a $5^\circ$ cut to retain the maximum
effect of regeneration at low energies.

The normalization accounts for the lack of knowledge in the value of the DM elastic scattering cross section, and to a lesser extent the
oscillation parameters. After normalization, see Fig.~\ref{fig:7}, the separation between the $\nu_\tau$ channel and the $\nu_e,~\nu_\mu$
channels remains and is still larger than the statistical error of the background and signal together. Using the regeneration effect, the
IC/DC effective volume could reasonably yield a determination of the neutrino flavor after 10 full years of operation.

%%%%%%%%%%%%%%%%%%%%
%%%%%%%%%%%%%%%%%%%%
\section{Conclusion}

We have investigated prospects of determining DM annihilation final states with neutrino telescopes by using the spectrum of
contained muon tracks from conversion of neutrinos that are produced in the annihilation of DM particles trapped inside the sun. Our
focus was on distinguishing neutrino final states from gauge boson and tau final states and on discriminating the
flavor of final state neutrinos. Gauge boson and tau final states are typically the dominant annihilation channels in supersymmetric models
with neutralino DM, while direct annihilation into neutrinos can occur in models that connect DM to the neutrino sector.
The theoretical motivation for the latter could provide the grounds for a dedicated analysis by the IceCube
Collaboration to put stringent bounds on annihilation to primary neutrinos, similar to what has been done for annihilation to gauge
bosons~\cite{Collaboration:2011ec}.

Primary neutrinos from DM annihilation result in a distinct peak in the muon spectrum at the DM mass. For DM
masses below $300$ GeV we can expect that the peak will be accessible to a detector the size of IC. The spectrum
is smeared as a result of the experimental error in energy reconstruction, but primary neutrinos may be distinguished from gauge boson
and tau final states after this effect is taken into account. We showed that for an energy resolution of 10 GeV (as in IC/DC) and by making
an optimal angular cut on the muons (which we found to be $5^{\circ}$), the branching ratios may be determined in the
parameter space within the reach of the one-year sensitivity limits of IC/DC with a km$^{3}$/yr of data. This is roughly
equivalent to $10$ years of data for a detector with the same capabilities and effective volume as IC/DC.

The regeneration of $\nu_\tau$ inside the sun may be used to distinguish the flavor of final state primary neutrinos. This effect becomes
important for DM masses above $300$ GeV and populates the spectrum with muons whose energy is well below
the energy of primary neutrinos. For DM masses up to about $500$ GeV, oscillations mix $\nu_\mu$ and $\nu_\tau$
effectively, which implies that regeneration affects final states with $\nu_\mu$ and $\nu_\tau$ similarly. Final states with $\nu_e$ are
therefore distinguishable within this mass range. For heavier DM particles, the $\nu_\mu-\nu_\tau$ oscillation becomes inefficient. As a
result, $\nu_\tau$ final states are picked out by the regeneration effect for DM masses
above $500$ GeV. We showed that final states with $\nu_\tau$ stand out at a statistically significant level for DM masses as heavy as
$800$ GeV, even after normalizing the muon spectrum to the total event count (to account for the unknown DM-nucleon elastic
scattering cross section). Again, such a distinction may be achieved with $10$ years of data from IC/DC.

In summary, using the the IC/DC results in tandem with independent measurements of the DM mass (for example, from the LHC), will
allow us to identify the annihilation channels of DM with multi-year data. Improved energy resolution and increased effective
volume of the detector will greatly help in achieving this goal.

%%%%%%%%%%%%%%%%%%%%%%%%%%%%%%%%%%%%%%%%%%%%%%%%%%%%%%%%
\section{Acknowledgements}

The authors wish to thank Bernard Becker, Tyce DeYoung, Arman Esamili, Yasaman Farzan, Alexander Friedland, Francis Halzen, Spencer Klein,
Robert Lauer, Danny Marfatia, Irina Mocioiu, and Carsten Rott for valuable discussions.

%%%%%%%%%%%%%%%%%%%%%%%%%%%%%%%%%%%%%%%%%%%%%%%%%%%%%%%%%

%\newpage

%\newpage

\end{document}